\documentclass{ws-p8-50x6-00}

\begin{document}

\title{Dynamics and statistics of simple models with infinite-range attractive
interaction}

\author{Micka{\" e}l Antoni}

\address{Thermodynamique et mod\'elisation des milieux hors d'\'equilibre,
Laboratoire de Chimie des Syst\`emes Complexes,
Universit\'e d'Aix-Marseille III, 
Av. Escadrille Normandie-Niemen, 
F-13397 Marseille, France\\
E-mail: m.antoni@lom.u-3mrs.fr}

\author{Stefano Ruffo}

\address{Dipartimento di Energetica ``S. Stecco'',
Universit\'a di Firenze, via S. Marta 3, I-50139 Firenze, Italy\\
INFM, Unit\`a di Firenze, I-50125 Firenze, Italy\\
INFN, Sezione di Firenze, I-50125 Firenze, Italy\\
E-mail: ruffo@avanzi.de.unifi.it}  

\author{Alessandro Torcini}

\address{Istituto Nazionale di Fisica della Materia, 
Unit\`a di Firenze, Largo E. Fermi 2, I-50125 Firenze, Italy\\
E-mail: torcini@ino.it\\
URL: http://torcini.de.unifi.it/$\sim$torcini}


\maketitle

\abstracts{The treatment of long-range interacting systems 
(including Newtonian
self-gravitating systems) remains a challenging issue in statistical
mechanics. Due to the lack of extensivity, they present non-standard
effects like negative specific heats, which shows the inequivalence of 
statistical ensembles (namely, microcanonical and canonical) even
in the limit of infinite number of particles ($N \to \infty$).  
In this paper we review a series of results obtained for one
and two dimensional simple $N$-body dynamical models with 
infinite-range attractive interactions and without 
short distance singularities. The free energy of
both models can be exactly obtained in the canonical ensemble, while
information on the microcanonical ensemble and on the dynamical
evolution can be derived from direct numerical simulations with simple
$O(N)$ codes, which make use of mean-field variables. 
Both models show a phase transition from a low energy clustered 
phase to a high energy gaseous state, in analogy with the models 
introduced in the early 70's by Thirring and Hertel. 
The phase transition is second order for the 1D model, first order for the
2D model. Negative specific heat appears in both models near the phase
transition point, but while for the 2D model it is an equilibrium
phenomenon, in the 1D case it is typical of transient metastable states,
whose lifetime grows with $N$. For both models, in the presence
of a negative specific heat, a cluster of collapsed particles coexists
with a halo of higher energy particles which perform long correlated
flights, which lead to anomalous diffusion: the mean square 
displacement grows faster than linear with time. The dynamical
origin of this "superdiffusion" is however different in the two models,
being related to particle trapping and untrapping in the cluster
in 1D, while in 2D the channelling of particles in an egg-crate
effective potential is responsible of the effect.
Both models are Lyapunov unstable and the maximal Lyapunov
exponent $\lambda$ has a peak just in the region
preceeding the phase transition. Moreover, 
in the low energy limit $\lambda$ increases proportionally 
to the square root of the internal energy, while
in the high energy region it vanishes as $N^{-1/3}$.
Since the 1D model is explicitely constructed considering the
first modes of a Fourier expansion of a classical
one dimensional gravity potential, the large scale properties of this
model in the low-energy clustered phase resemble those of 
1D gravity (mass-sheet models). The relation of both models with gravity
remains to be analysed in detail by adding successive Fourier waves.
However, we believe that the main dynamical effects we have found:
superdiffusion and "strong" chaos (due to the evaporation and sticking
of particles to the cluster) should be present also in "true" gravitational
systems, and we see no obstruction that they appear also in 3D.
}

\section{Introduction}
It is well known that extensivity is an essential ingredient
for building up thermodynamics from statistical mechanics
questions (see e.g. Landau's textbook \cite{Landau}). One is 
therefore naturally led to ask questions about what happens 
when internal energy, entropy and other thermodynamic quantities 
are no more extensive, i.e. when a part of a system has not the
same thermodynamics properties of the whole.
This problem naturally arises in Newtonian gravity, where the
inherent long-range of the forces is the ultimate cause of 
non-extensivity,
but is also present in plasma physics although the coexistence of
attractive and repulsive long-range Coulomb interactions determines
the screening effect which mitigates this problem.
Therefore, one knows that statistical physics does not have
the same strong impact for gravitational systems as it has for systems 
with short-range interactions and hard-cores, although one would 
really like to use statistical concepts in this field\footnote{
Recently, Tsallis and co-workers have developed a new
interesting approach based on an alternative definition of entropy,
which could cope with non-extensivity\cite{tsallis}. The relation 
of their approach with our results remains to be carefully investigated}.

On the other hand mean-field models are frequently used in statistical
mechanics as a first simple tool to understand the collective behavior of
systems with short-range interactions and it is believed that they 
describe their "universal" features (such as critical exponents at phase
transitions) above a certain "critical" number of neighbors. 
In this context, it
is curious that none really posed the question of the validity of 
the statistical approach, considering that mean-field models (expecially 
those written in terms of an infinite-range Hamiltonian) 
explicitly violate extensivity.
A reason for that lies in the fact that the thermodynamic limit is performed
resorting to saddle-point techniques, which put the Hamiltonian
in the explicitly decoupled form, hiding the difficulties inherent in
the long-range interaction. However, all concepts related to statistical 
ensemble equivalence (like microcanonical/canonical equivalence) are open
for such systems, and we will review in this paper models of this type
for which it is indeed violated, producing effects like negative specific
heats within the microcanonical ensemble, which are well known in the
gravitational context~\cite{lind2}.

The central idea contained in the papers which we are reviewing here is 
that it is possible to establish a close correspondence between the behaviors
observed in the Hamiltonian dynamics of $N$-body self-gravitating systems
and those of much simpler models of the mean-field infinite-range class.
We have dealt only with systems in one and two spatial
dimensions, but this approach can in principle be extended also to three dimensions. 
One of these models has been indeed rigorously derived from
the self-gravitating model in one dimension, by restricting the line
to a finite segment and truncating the Fourier expansion of the potential 
to its first wave component~\cite{ant1}. The two-dimensional model
is less tightly related to two dimensional gravity, but arises from a 
similar truncation of the Fourier expansion in a box~\cite{tor1}. 
The truncation is not ineffective and introduces features 
which are not present in the original self-gravitating system. 
In one dimension, the truncated model displays a 
second order phase transition from a phase where the mass points
are grouped in a cluster to one where they are homogeneously dispersed;
this phase transition is not present in the full self-gravitating system, 
which is always in the clustered phase 
(being the potential confining). However, we 
claim that many features of the clustered phase of our mean-field 
model are similar to those observed in the full model. Moreover, our model
can be a toy system for studying phase transitions in the simple setting
of one dimension (with all the computational advantages), since it is 
expected that phase transitions are present in 
three dimensions~\cite{Chavanis}. In two dimensional models
some theoretical result \cite{abdalla} seem to
suggest that a phase transition occurs at finite temperature
from a clustered to a homogeneous phase.

The advantage of studying one and two-dimensional mean-field models
is that their free energy in the canonical ensemble can be exactly derived by
performing the mean-field limit (the infinite $N$ limit at fixed volume),
which is a reasonable surrogate of the thermodynamic limit 
(the infinite $N$ limit at fixed density).
Unfortunately, at variance with Thirring models
~\cite{her1,her2}, the microcanonical solution (for instance the entropy-energy 
relation) is not yet available for such models (although
some attempts already exist along this line~\cite{ant3}). However,
plenty of informations can be gotten from relatively fast 
numerical simulations, thanks to the algorithmic advantage of 
working with ${\cal O}(N)$ codes (instead of usual ${\cal O}(N^2)$), 
due to 
the introduction of mean-field variables. Moreover, important informations
can be obtained from solving the one-dimensional non collisional
Boltzmann-Poisson (BP) equation~\cite{Ina}, although it corresponds to
inverting the $N \to \infty$ limit with the infinite time limit, which
can be a dangerous exchange in the presence of transient states.

In particular, from the analysis of the BP equation for the 1D 
model \cite{Ruffo},
it turns out that all the initial conditions with a homogeneous
particle distribution and with a symmetric
velocity distribution, are stable in time.
Moreover, extensive dynamical simulations of the 1d and 
2D model allowed us to study in detail aspects
like non-ergodicity and long-time memory,
that are expected phenomena 
in systems with non-screened long-range interactions
\cite{prigo}. 

The structure of this paper is the following. After introducing and discussing
the main models in Section 2, we derive statistical equilibrium properties in 
Section 3. In this Section we comment also on the presence of metastable
states near the phase transition point in the one-dimensional model; these
states have a lifetime which grows with $N$. In the two-dimensional model,
the microcanonical simulations reveal a region of negative specific heat.
Dynamical properties are discussed in Section 4. The main effect is anomalous
diffusion, superdiffusion in our case, which means that the mean square 
displacement of a single point mass grows in time faster than linear. In the 
one-dimensional model this phenomenon appears as a consequence 
of the presence of metastable states, i.e. it is a transient-to-equilibrium 
effect and is generated by the correlated trapping and untrapping of the 
particles in the cluster.
In the two-dimensional model it is instead due to the generation of
an effective mean-field potential of the egg-crate form, where particles
in a certain energy range can perform long free flights; the phenomenon
is destroyed by noise on the long-time span, but noise dyes out as
$N \to \infty$ and thus superdiffusion extends to the time asymptotic
regime. In Section 5 it is shown that the dynamics of both models is
chaotic, by computing the maximal Lyapunov exponents $\lambda$. 
Since the systems are integrable for vanishing internal
energy $U$ and in the high energy limit, $\lambda$ 
should exhibit a peak at intermediate energy. Indeed,
$\lambda$ is maximal in the phase
transition region.  In the high energy regime $\lambda$ 
vanishes with an universal scaling
law $\lambda=\lambda(N) \sim N^{-1/3}$, while at low energy it 
is positive also in the mean-field limit and
increases as $\lambda=\lambda(U) \sim U^{1/2}$. 
In Section 6 we finally draw some conclusions 
and discuss some perspectives of this novel approach.

\section{The Models}\label{sec:mod}

In this Section we introduce the 1D and 2D models. 
We discuss their Hamiltonians and equations of motion
by introducing the mean-field variables. We also give
a brief description of the phenomenology and of the
tools that have been introduced for their study.

\subsection{One dimensional model}\label{subsec:1d}

\noindent

This first model describes a collection of $N$ identical particles that move 
under their mutual gravitational-like attraction. Each particle is a
mass point and boundary conditions are periodic. Particles
are therefore constrained to move on the unit circle, {\it i. e.} in a one 
dimensional geometry.  The total energy is constant 
and given by~\cite{ant1}:
\begin{equation}
H = \sum_{i=1}^N \frac {p_i^2} {2 m} 
+ \frac {m^2} {2N} \sum_{i,j=1}^N \Bigl(1-\cos(\theta_i-\theta_j)\Bigr)=K+V,
\label{1dham}
\end{equation}
where $\theta_i \in ]-\pi,\pi]$ is the coordinate of the
$i$-particle, $p_i$ its 
conjugate momentum and $m$ the mass of the particles.
A time-discrete version of this model has been previously introduced
by Kaneko and Konishi~\cite{kan1}.
The system is isolated, total momentum is conserved and can be set to 
zero without loss of generality. 
$K$ and $V$ are the kinetic and potential energy respectively. 
As the latter has no singularity when $i=j$, the particles do 
not collide but smoothly cross through one other. This shows up 
clearly in the equations of motion

\begin{equation}
\ddot \theta_i = \frac {\dot p_i} m = - \frac {m} N \sum_{j=1}^N
\sin(\theta_i-\theta_j)~,
\label{eqmv}
\end{equation}
since they correspond to a system of fully coupled pendula.

The potential energy is rescaled by $1/N$ in order to get a finite specific 
energy ${\cal H}=H/N$ in the thermodynamic limit $N \to \infty$, hence 
the potential is thermodynamically stable ($H \ge E_0 N$).
The characteristic period $t_c$ of particle motion is then
an intensive quantity, given by $t_c=2\pi/\sqrt{m}$. In the large $N$ 
limit, particles can be treated in a mean-field approximation and the 
distribution functions can be computed using Poisson-Boltzmann 
techniques~\cite{Ina,lat1}. 
These techniques have been used in a large variety of dynamical systems 
(plasmas, galaxies, fluids) where particle-particle collisions can be neglected 
when compared to the particle-mean-field coupling. For model (\ref{1dham}), the 
mean-field variable writes: 

\begin{equation}
{\bf M}=(M_x,M_y)=\frac 1 N \sum_{j=1}^N {\bf s}_j \quad, 
\quad {\rm with} \quad {\bf s}_j = (\cos \theta_j, \sin \theta_j) \quad.
\label{meanfield}
\end{equation}
With these definitions, Eq. (\ref{eqmv}) reduces to that of a 
single perturbed pendulum

\begin{equation}
\ddot \theta_i = - M \sin(\theta_i-\phi)
\label{eqmv1}
\end{equation}
where we have set the mass $m=1$ for simplicity
and $\phi={\rm arg}({\bf \rm M})$ is the phase of the mean-field. 
Both $M$ and $\phi$ are scalar 
quantities that depend on time through $\theta's$ and their time 
evolution describes the collective dynamics of the system.

The single-particle dynamics is ruled by the Hamiltonian
\begin{equation}
h_i = \frac{p_i^2}{2} + 1- M_x \cos (\theta_i) - M_y \sin (\theta_i) 
= K_i + V_i
\label{single}
\end{equation}
which is non authonomous, since it depends on the mean-field, which reflects
the collective dynamics of the model.  Hence, each particle moves 
in a mean-field potential $V_i$ determined by the instantaneous 
positions of all the other particles of the system.

The thermodynamics of model (\ref{1dham}) can be derived theoretically in
the canonical ensemble (see Sect. 3). 
It shows a second-order phase transition at a
critical value of the internal energy $U_c=H_c/N=0.75$ between 
a collapsed phase, 
where a single cluster is present, and a homogeneous phase, where particles 
are uniformly distributed on the circle. The development of the cluster in 
the collapsed phase corresponds to the well known Jeans instability for 
self-gravitating systems~\cite{Ina}.

Numerical simulations of self-gravitating systems in their collapsed phase
and initially far from equilibrium, usually reveal two evolution 
mechanisms: i) collective phenomena that involve almost all the particles 
and that rapidly drive the system into a quasi-stationary configuration;
ii) individual effects that rule the successive slow relaxation to 
equilibrium. On short time scales, spiral structures are often developed 
to eliminate the excess of kinetic energy~\cite{ant1}; they are due 
to the formation phase-space domains with an excess or depletion of 
particles. These structures are robust enough to 
resist individual particle effects and are known to prevent the 
system from reaching equilibrium~\cite{rou1}. They are similar to the
Dupree structures observed in plasmas where they inhibit Landau damping
and constitute an important element of Vlasov turbulence~\cite{ant2}. 

Model (\ref{1dham}) contains only one spatial scale, the length 
$2 \pi$; this reflects in the presence of a single modulation of
the particle density (the cluster). However, this model belongs to the family of 
$N$-body Hamiltonian systems that write

\begin{equation}
H_s =\sum_{i=1}^N \frac {p_i^2} {2 m} + \frac {m^2}{2N}
\sum_{i,j=1}^N \sum_{n=0}^s \frac 1 {k_n^2}
\Biggl(1-\cos\Bigl[k_n(\theta_i-\theta_j)\bigr]\Biggr),
\label{ham2}
\end{equation}
where $k_n=2n+1$ and where the parameter $s$ counts the number of Fourier 
harmonics~\cite{ant1,Els}. The interest in this family of Hamiltonians 
lies on the fact that in the limit $s \to \infty$ they write
 
\begin{equation}
H_{\infty} = \sum_{i=1}^N \frac {p_i^2} {2 m} 
+ \frac {m^2}{2N} \sum_{i,j=1}^N |\theta_i-\theta_j|= 
\sum_{i=1}^N \frac {p_i^2} {2 m} + 2 \pi G \frac {m^2}{N} 
\sum_{i,j>i} |\theta_i-\theta_j|,
\label{ham3}
\end{equation}
where $G$ is the gravitational constant and the customary
dimensionless units $1=2 \pi G$ have been chosen~\cite{mil1}. Up to 
the $1/N$ scaling constant of the potential, this Hamiltonian is the one 
of a collection of $N$ identical planar mass sheets of 
constant mass surface density $\mu$ evolving in one dimension perpendicular to 
their surface. This model was originally introduced to describe the dynamics 
of stars evolving in the neighborhood of a flattened galaxy~\cite{oort1}.
For the Poisson-Boltmann limit to exist, one must
require the the total mass $\mu N = m$ remains finite when one lets 
$N \to \infty$. Then
\begin{equation}
{\cal H}_{\infty} = \frac {H_{\infty}}{N} = \sum_{i=1}^N \frac 
{{\bar p}_i^2} {2 \mu} 
+ 2 \pi G \mu^2 \sum_{i,j>i} |\theta_i-\theta_j|,
\label{hamsheet}
\end{equation}
where ${\bar p}_i = p_i / N$ are the scaled momenta.  
We would like to notice that the Hamiltonian $H_\infty$
can be transformed in that corresponding to the usual 
gravitational 1D model also
by simply multipling the time units by a factor $\sqrt{N}$,
this obviously corresponds also to rescale the energy units by
a factor $N$ \cite{Anteneodo}.

The dynamics of the mass sheet model involves an interplay of all the 
spatial scales, which is the source of its metastability. The advantage 
of using models (\ref{ham2}) is that they allow a selection of a limited
number of modes, through the tuning of the parameter $s$; the qualitative
changes in the dynamical and  in the thermodynamical behavior can be followed 
by adding higher Fourier components.  Such a multiscale approach is based 
on the mean-fields ${\bf M}_{k_n}=1/N \sum_j {\bf s}_{j,k_n}$ with 
${\bf s}_{j,k_n}=(\cos k_n\theta_j,\sin k_n \theta_j)$.
This approach to self-gravitating systems is similar to the one proposed in 
Ref.~\cite{kies1} where a weak formulation based on a smoothed-out 
gravitational 
potential is introduced. We expect that the large scale
properties won't be strongly modified by the introduction of 
smaller spatial scales. 
However, the balance between collective (large scale) and 
individual (small scale) effects might depend crucially on the
number of Fourier components.
The investigation of the properties of model (\ref{1dham}) is thus 
the very first step in the investigation of Hamiltonian (\ref{hamsheet}). 
Indeed, one major difference is that while model (\ref{1dham}) has
a second-order phase transitions, model (\ref{hamsheet}) has no
phase transition.
The absence of phase transitions is a trivial consequence of the existence
of the scale transformation $\theta_i \to \alpha^2 \theta_i$,
$p_i \to \alpha p_i$, $t \to t/\alpha$, which transforms the
energy as ${\cal H}_{\infty} \to \alpha^2 {\cal H}_{\infty}$
and leaves the equations of motion unchanged. Thus, if $\alpha > 1$
states corresponding to a larger energy coincide with those
at a shorter time. Therefore, increasing the energy cannot produce
any phase transition, but it is like observing the dynamics on a
shorter time scale. 
Hamiltonian (\ref{hamsheet}) is always in the clustered phase, since
the potential is always confining. To induce a phase transition one must
explicitly introduce a scale in the model (this has been recently
done for the spherically concentric mass shell model in 
Ref.~\cite{mill1}).
An open question is how the phase transition disappears in the
$s \to \infty$ limit, and a related one is whether phase transitions
are present in 2D and 3D self-gravitating systems~\cite{Chavanis}.

\subsection{Two dimensional model}\label{subsec:2d}

We have also studied a two dimensional model, that can be viewed
as a generalization of the one dimensional model introduced in the 
previous Subsection. We consider a system of $N$ pointlike
identical particles with unitary mass moving in a $2$-D 
box with periodic boundary coditions.
The dynamics is ruled by the Hamiltonian
\begin{eqnarray}
&& H = \sum_{i=1}^N \frac{p_{i,x}^2+p_{i,y}^2}{2}
\nonumber \\
&& +\frac {1}{2N} \sum_{i,j}^N
\Biggl[3- \cos(x_i-x_j)
-\cos(y_i-y_j)-\cos(x_i-x_j)\cos(y_i-y_j)\Biggr] 
\nonumber \\
&&= K+V \quad ,
\label{2dham}
\end{eqnarray}
where $(x_i,y_i) \in ]-\pi,\pi] \times ]-\pi,\pi]$,
$(x_i,p_{i,x})$ and $(y_i,p_{i,y})$ are pairs of conjugate
variables. $K$ (resp. $V$) is the kinetic (resp. potential) energy. 
The  reference energy is chosen in order to ensure
the vanishing of the total energy of the system when all 
the particles are placed in the same position ($V=0$) at rest ($K=0$). 
Without the presence of the third term in the potential the particles 
would move independently
along the $x$ and $y$-directions according to the 
previously introduced $1$-D potential (\ref{1dham}).
The interparticle potential appearing in (\ref{2dham}) 
mimicks the Fourier expansion in a periodic square box 
of side $L=2 \pi$ of a $2$-D self-gravitating Newtonian potential 
($\log |r|$, being $r$ the interparticle Euclidean distance), 
restricted to its first three terms $(n_x,n_y) = (0,1), (1,0), (1,1)$,
where $k_x= 2 \pi n_x/ L$ and $k_y= 2 \pi n_y/L$ are the wave
numbers along the two directions.
Interactions of the  $\log|r|$ type arises also for point 
vortices in  2D Eulerian turbulence~\cite{Nov}, and our approach
could be fruitfully extended to this case. 

The equations of motion for the coordinates $x_i$ are
\begin{equation}
\ddot x_i= -\frac 1 N
\sum_{j=1}^N\Biggl[\sin(x_i-x_j)+\sin(x_i-x_j)\cos(y_i-y_j)\Biggr] 
\label{2deqm}
\end{equation}
and those for $y_i$ are obtained exchanging $x \leftrightarrow y$,
due to the symmetry of (\ref{2dham}).

Previous investigations of model (\ref{2dham})~\cite{tor1}
have revealed that at low energy $U=H/N$ the particles,
due to attractive coupling, collapse into a unique cluster, with 
$U$-dependent spatial extension. This clustered phase 
survives up to a critical energy $U_c \sim 2.0$
(numerically determined from an implicit equation derived
in the canonical ensemble), where a first-order phase transition 
occurs to a homogeneous phase. 
This transition will be discussed in more detail in the next 
Section. The collective behaviour is evidenced by
the following mean-field variables
\begin{equation}
{\bf M}_1 = (<\cos(x)>_N,<\sin(x)>_N) 
= M_1 \enskip ( \cos(\phi_1),\sin(\phi_1) )
\label{meanfield1}
\end{equation}
\begin{equation}
{\bf M}_2 = (<\cos(y)>_N,<\sin(y)>_N) 
= M_2 \enskip ( \cos(\phi_2),\sin(\phi_2) )
\label{meanfield2}
\end{equation}
\begin{equation}
{\bf P}_1 = (<\cos(x + y)>_N, <\sin(x +  y)>_N) 
= P_1 \enskip ( \cos(\psi_1),\sin(\psi_1) )
\label{meanfield3}
\end{equation}
\begin{equation}
{\bf P}_2 = (<\cos(x - y)>_N, <\sin(x - y)>_N) 
= P_2 \enskip
( \cos(\psi_2),\sin(\psi_2) )
\label{meanfield4}
\end{equation}
where $<..>_N$ denotes the average over all the particles in
the system. 
The moduli $M_{1,2}$ and $P_{1,2}$ are maximal and equal to 1
when all the particles have the same position and
their value decrease when the spatial distribution of the 
particles extends. For a homogeneous distribution, according
to the central limit theorem, $M_{1,2} \approx P_{1,2} \approx 
O(1/\sqrt{N})$. These quantities are the
order parameters characterizing the degree of clustering 
of the system.

By reexpressing the equation of motion (\ref{2deqm}) for both 
coordinates $x$ and $y$ in terms of the mean-field variables, one 
straighforwardly shows that the time evolution of the $i-$th particle 
is ruled in a way similar to the 1D model
by the following single particle non autonomous Hamiltonian :
\begin{eqnarray}
h_i&& = \frac{p_{x,i}^2+p_{y,i}^2} 2 
+ \Bigl[3-M_1\cos(x_i-\phi_1)-
M_2\cos(y_i-\phi_2)
\nonumber \\
&&-\frac 1 2 P_1\cos(x_i+y_i-\psi_1)-
\frac 1 2 P_2\cos(x_i-y_i-\psi_2) \Bigr] 
\nonumber \\
&&= K_i + V_i   \quad   .
\label{2dham1}
\end{eqnarray}
\noindent
Since $V$ is invariant under the transformations 
$x \leftrightarrow -x$, $y \leftrightarrow -y$ and 
$x \leftrightarrow y$, it turns out that in the 
mean-field limit ($N\to \infty$ with constant
$U=H/N$), $M_1=M_2=M$ and $P_1=P_2=P$. 

\section{Equilibrium properties}\label{sec:equilib}

In this Section we will mainly discuss the canonical equilibrium
exact solutions of the previoulsy introduced 1D~(\ref{1dham}) and 
2D~(\ref{2dham}) self-gravitating models and we will compare these
solutions with the result of the microcanonical simulations. 
The canonical free energy is easily
obtained in both cases by resorting to an inverse Gaussian transformation 
(sometimes called the Hubbard-Stratonovich trick) and evaluating
an integral with saddle-point techniques. For the 1D model~(\ref{1dham}), 
we have shown 
that the phase transition between the collapsed and the homogeneous 
phase is a second order phase transition. For the 2D model~(\ref{2dham})
the transition is instead of first order. Microcanonical simulations
reveal the presence of a negative specific heat region near the phase
transition for both the 1D and the 2D model, but for the former case
this is only a transient effect (however, the transient grows with
the number $N$ of point masses).
Ensemble inequivalence and negative specific heats are indeed rather 
familiar to astrophysicts and appear in the phenomenon known
as the {\it gravothermal catastrophe}~\cite{lind1}.
These observations confirm the conjecture of 
Lynden-Bell~\cite{lind1,lind2} 
and demontrate on a simple model that the 
equivalence among statistical ensembles is violated due
to the long-range nature of the forces and to the absence
of repulsive hard-cores.

\subsection{One dimensional model}\label{subsec:1d_eq}

\begin{figure}[h]
\centerline{\psfig{figure=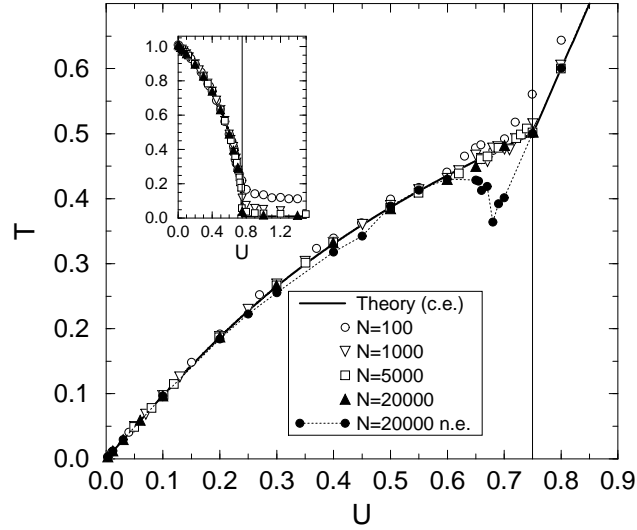,width=10cm}}
\caption{Temperature $T$ versus the internal energy $U$ for the 
1D model (\protect\ref{1dham}). The full curve is obtained
theoretically in the canonical ensemble ($N \to \infty$).
The points are the result of microcanonical simulations
with an increasing number of mass points, in particular 
the full circles refer to an initial "water bag distribution".
In the inset, the mean-field variable $M$ vs. U is reported
with the same notation. The horizontal lines indicates 
$U_c$.
\label{f1}
}
\end{figure}

The one-dimensional model of Eq.~(\ref{1dham}) can be exactly
solved in the canonical ensemble. Its free energy 
\begin{equation}
F=- \lim_{N \to \infty} \frac{1}{N} 
\ln \int \prod_i d \theta_i d p_i
\exp (- H/T)
\end{equation}
is given by the expression
\begin{equation}
-F/T= \frac{1}{2} \ln (2 \pi T) - \frac{m}{2T} +
\mbox{max}_x \{ \frac{-x^2 T}{2m} + \ln (2 \pi I_0(x)) \}
\end{equation}
where the $\mbox{max}$ is obtained by solving the consistency
equation
\begin{equation}
\frac{I_1(x)}{I_0(x)}= \frac{T x}{m}~,
\label{cons}
\end{equation}
where $I_0$ and $I_1$ are respectively the zero and first order
modified Bessel functions. In the following we will adopt the
value $m=1$, without loss of generality.
Eq.~(\ref{cons}) has the unique solution $x=0$, corresponding to the
vanishing mean-field region $M=0$, for $T \ge 0.5$; it
has instead two symmetric nonvanishing solutions, corresponding
to a nonvanishing value of $M$ for $T < 0.5$. This latter is the 
clustered phase and the ratio $M/T$ determines the average size
of the cluster (which becomes infinitely narrow, i.e.
it collapses to a point, only in the zero temperature limit).
The temperature dependence of the mean field $M$ is exactly derivable 
and it is shown in the inset of Fig.~\ref{f1} by the continuous line. 
The temperature-energy 
relation can be also derived using the standard formula 
$U (T) =\partial (F/T)/\partial (1/T)$ and is reported in the same figure.
This model has therefore a second order phase transition at the critical
temperature $T_c=0.5$, corresponding to the critical energy $U_c=3/4$.
The dynamical time averages of the same quantities can be determined
by computer simulations. Being the model Hamiltonian, the energy is
fixed by the initial state and temperature is dynamically determined
by computing the average kinetic energy per particle, and waiting
for long time relaxation.

\begin{figure}[h]
\centerline{\psfig{figure=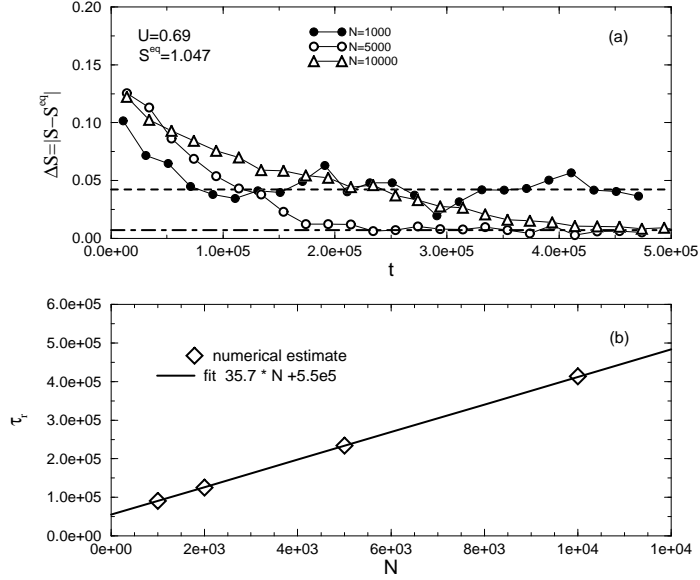,width=10cm}}
\caption{a) Absolute difference with respect to the equilibrium
Boltzmann entropy for the momentum distribution vs. time $t$. 
Relaxation slows down for increasing $N$.
b) A plot of the relaxation time $\tau_p$ vs. the number of
particles $N$. The full line is a best linear fit to the data.
\label{f6}
}
\end{figure}

In all energy ranges, except close to the critical point, every chosen
initial state relaxes quite fast to the canonical curves of $M$ vs. $U$
and $T$ vs. $U$; finite $N$ corrections are visible in Fig.~\ref{f1}.
On the contrary, in a region just below the critical point (evidenciated
by the vertical line in Fig.~\ref{f1}) initial states of the "water-bag" class
(uniformly distributed momenta in a fixed interval symmetric around
zero and all positions at $\theta_i=0$) show a relaxation to metastable
non-equilibrium states, whose $T$ vs. $U$ curve has two main
features: i) for $U > 0.67$ the points lye on the continuation for
$U<3/4$ of the line $T=2U-1$; the corresponding states are gas-like,
with zero magnetization (uniform in space) and have non-Gaussian tails
in momentum; ii) for $0.6<U<0.67$, $T$ decreases when $U$ increases,
this is the negative specific heat region. All these latter points, obtained
for a system of $N=20,000$ point masses, relax very slowly to the
canonical curve (as shown by the behavior of their relative Boltzmann
entropy in Fig.~\ref{f6}a) in a time which increases linearly with the
number $N$ of particles of the system Fig.~\ref{f6}b.
Ideally, if one would have prepared a system with an infinite number
of particles, this would not have relaxed to the canonical equilibrium.
The metastability of our model is not as strong as the one encountered
for disordered systems (where the relaxation time diverges
exponentially in $N$), it is of the same type as the one found for
the mass-sheet model and for plasma models as reviewed in Ref.
\cite{Els}. Its origin lies in the complex structure of the 
chaotic web through which the orbit channels in phase-space
(see further comments Section 4.1).

The equilibrium behavior discussed above and depicted in Fig.~\ref{f1}
can be derived also using the non-collisional Boltzmann-Poisson (BP)
equation in two different ways: i) imposing the maximal entropy
principle, as done in Ref.~\cite{Ina}; ii) looking for solutions with 
a Gaussian shape in momentum and a factorized stationary distribution
function $f(\theta,p,t)=f_1(\theta)f_2(p)$~\cite{lat1}.
To our knowledge, the fact that the canonical equilibrium solution
coincides with the one given by the BP equation is a novel feature,
which deserves further investigations for other models. The use of the
BP equation gives access also to the one-particle distribution
function $f(\theta,p,t)$, on which numerical checks have been
performed in Ref.\cite{lat1}. A feature which remains unexplored in the 
context of the PB equation is metastability, but recently one of the
authors~\cite{Ruffo} has proven, using the expression of the evolution
operator in terms of Lie brackets~\cite{Perez}, that initial states
symmetric in momentum $f_1(p)=f_1(-p)$ and uniform in position
do not evolve in time to all orders in perturbation theory. This is a first
sign of metastability in the PB context, where since one is treating 
directly the $N\to \infty$ limit, metastability manifests itself in
a multi-stability. Finally, let us mention that recently 
in Ref. \cite{casetti} it has been shown that
the transition observed for the model  (\ref{1dham}) 
is associated, in the mean-field limit,
to a change in the topology of the corresponding 
configuration space.

\subsection{Two dimensional model}\label{subsec:2d_eq}

\begin{figure}[h]
\centerline{\psfig{figure=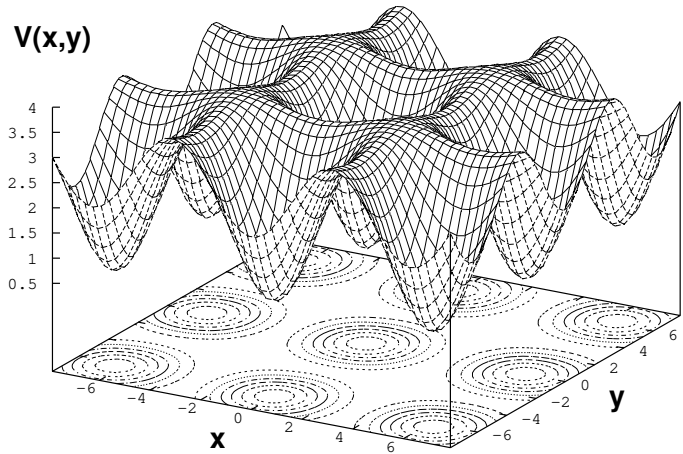,width=8cm,angle=270}}
\caption{Instantaneous single particle potential $V(x,y)$ together
with its contour-plot for $U=1.00$ and $(x,y) \in [-5 \pi/2, 5 \pi/2]^2$.
\label{f2}
}
\end{figure}

\begin{figure}[h]
\centerline{\psfig{figure=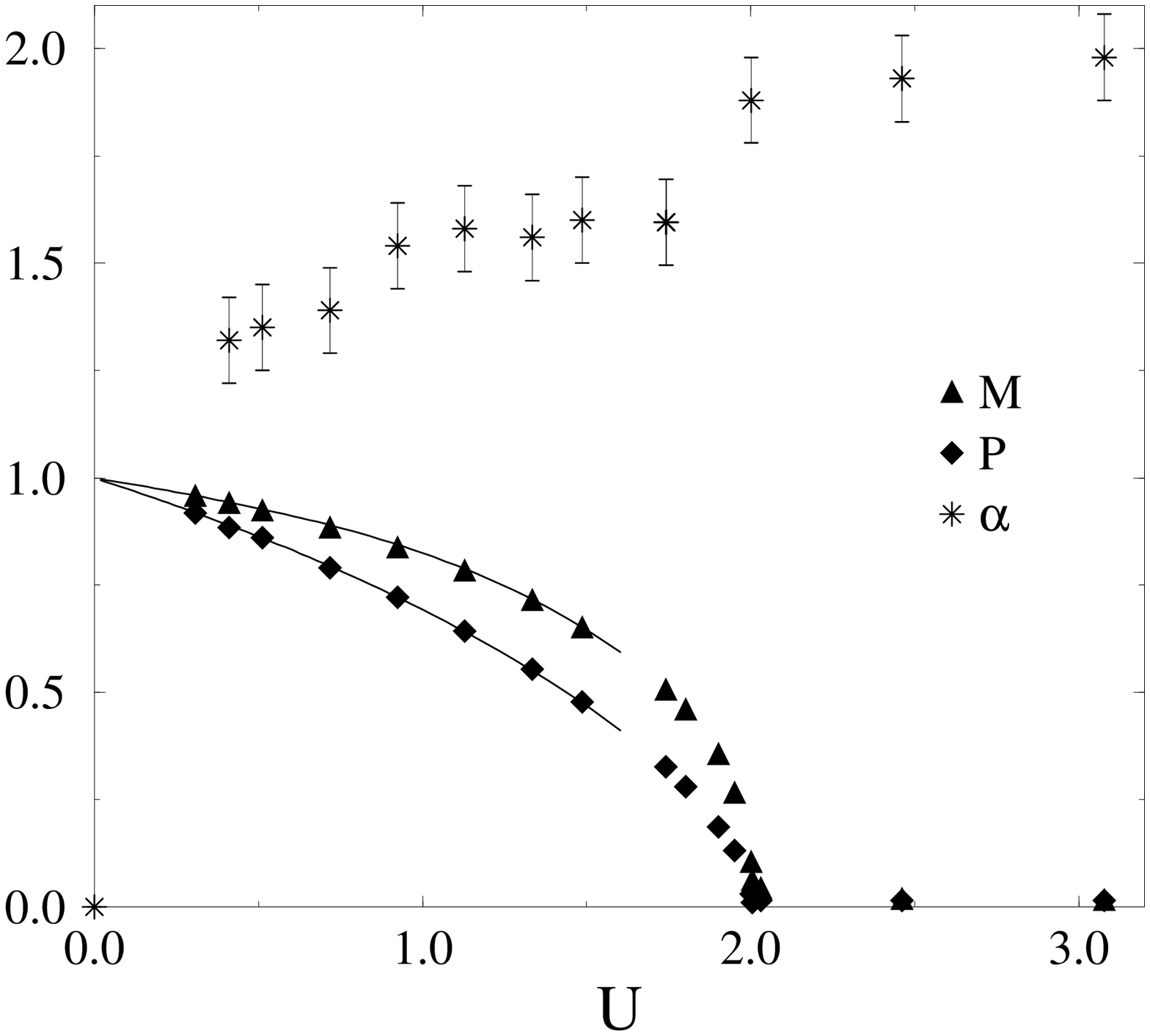,width=8cm,angle=270}}
\caption{Meanfield quantities $M$ and $P$ as a function of $U$.
The solid line refer to canonical results analytically estimated,
while the symbols to microcanonical results obtained via numerical
simulations. The exponent $\alpha$ defined in eq.
(\protect\ref{msqd}), are also shown (asteriskes).
The data have been obtained with $N=4,000$ (apart 
few points with $N=10,000$) and averaged over a 
total integration time of order $10^6$.
\label{f3}
}
\end{figure}

Model (\ref{2dham}) can be solved in the canonical ensemble
as in the 1D and all its equilibrium features can be exactly derived
in the mean-field limit, using again saddle-point technique
\cite{tor1}. We do not reporte here the explicit expression
for the free energy, that is already reported elsewhere~\cite{tor1}.
The single particle potential $V_i$ for the 2D case
(\ref{2dham1}) is reported in Fig.~\ref{f2}, 
it has a shape similar to the so-called 
egg-crate potential studied in Ref.~\cite{geis}.
The potential is periodic along the two spatial
directions and in each elementary cell has 4 maxima 
($V_M=3+2M-P$), 4 saddle points ($V_s=3+P$) and a
minimum ($V_m=3-2M-P$). The depth of the potential well
is $(V_s-V_m)$ and the center of the cluster in the collapsed
phase coincide with the position of the minimum of the
potential.  In the limit $U \to 0$,
$M$ and $P \to 1$ and therefore $V_M \to 4$,
$V_s \to 3$ and $V_m \to 0$. In this limiting
situation all the particles are trapped 
in the self-consistent potential well.
Increasing $U$, the kinetic energy
produces an evaporation of the particles from the cluster.
This implies that also the values of $M$, $P$ and of the well depth 
decrease (see Fig.~\ref{f3}). For $U \ge U_c$, there are no more 
clustered particles and in the limit $N \to \infty$ the single 
particle potential becomes flat $V_M = V_s = V_m = 3$ and 
time independent.

However, due to finite $N$ effects, the instantaneous mean-field 
variables fluctuate with a typical time ${\cal O}1$ within the statistical 
band range $\sim 1/\sqrt{N}$. 
This implies that in the collapsed phase the particles move 
in a fluctuating potential.

\begin{figure}[h]
\centerline{\psfig{figure=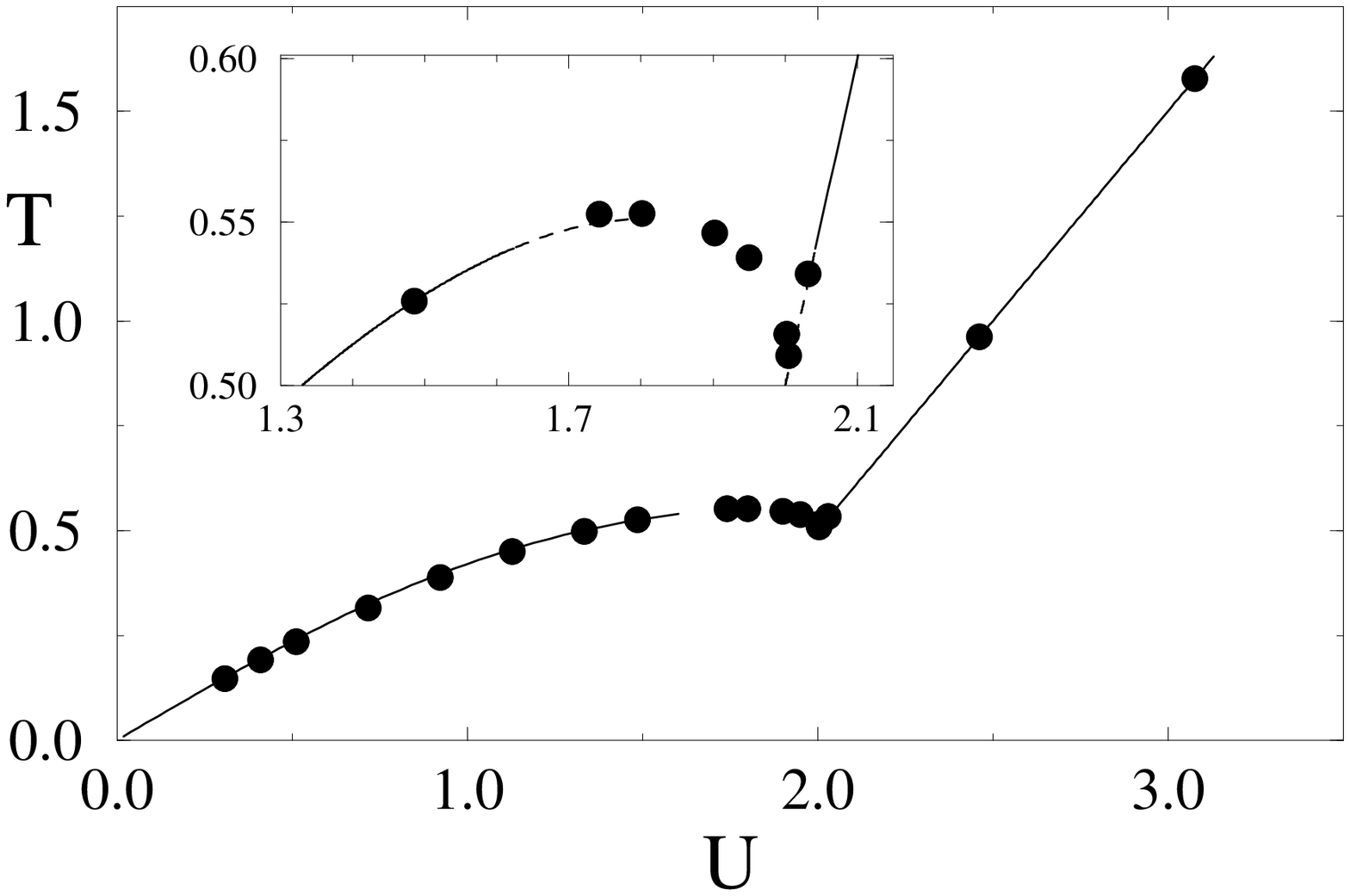,width=6.5cm,angle=270}}
\caption{Temperature $T$ as a function of the internal energy $U$
for the 2D model. The solid line refers to analytical
results obtained within the canonical ensemble, while the full
circles to microcanonical simulation results.
The inset is an enlargement of the transition region:
the full curves (resp. dashed) refer to the absolute 
(resp. relative) minimum of the free energy $F(T)$. 
\label{f4}
}
\end{figure}

In Fig.\ref{f4} the temperature $T$  is reported as
a function of $U$. Above $U_c \sim 2.0$, $T$ increases linearly 
with $U$
indicating that the system behaves like a free particle
gas. In the collapsed phase, the tendency of the system to collapse
is balanced by the increase of the kinetic energy
\cite{compagn}. This competition
leads initially (for $ 0 < U < 1.8$) to a steady
increase of $T$, followed (for $ 1.8 < U < U_c$)
by a rapid decay of $T$. This yields a negative
specific heat as illustrated in the inset
of Fig.\ref{f4}. These results are in full
agreement with theoretical predictions based on the
analysis of a simple classical cell model \cite{her2} and
with numerical findings \cite{compagn,posch} for short
ranged attractive potentials without hard-cores.
The phenomenon of negative specific heat can be
explained within a microcanonical approach with
an heuristic argument \cite{her2}: approaching
the transition, a small increase of $U$ leads
to a significative reduction of the number of collapsed
particles (as confirmed from the drop exhibited
by $M$ and $P$ for $U > 1.8$ in Fig.\ref{f3}); 
as a consequence
the value of $V$ grows and, due to
energy conservation, the system becomes cooler.

Our data confirm also another important prediction of
Hertel and Thirring \cite{her1,her2}: the non-equivalence of
canonical and microcanonical ensemble nearby the transition region.
In the inset of Fig.~\ref{f4} we report the microcanonical
findings, obtained by direct simulations,
and the theoretical canonical results. These two sets of
data coincide everywhere, except in the energy
interval $1.6 \le U \le 2.0$.
The discrepancy is due to the impossibility of the canonical ensemble
to exhibit a negative specific heat, a prohibition that does not hold for
the microcanonical ensemble. Our theoretical estimation
of the Helmholtz free energy $F=F(T)$ reveals that usually
$F$ has an unique minimum. For $T < 0.5$ the minimum $F_{c}$
corresponds to non zero values of $M$ and $P$ (i.e. to the collapsed
phase), while for $T > 0.55$ the minimum $F_{H}$ is associated
to $M=P=0$ (i.e. to the homogeneous phase). In the region $0.5 < T < 0.55$,
both minima $F_c$ and $F_{H}$ coexist as local minima
of the free energy. However, for $T < T_c = 0.54$
the collapsed phase is observed because $F_c < F_{H}$ , while
for $T > T_c$ the homogeneous phase prevails since $F_{H} < F_c$.
At $T=T_c$ the two minima are equivalent
and a jump in energy from $U(T_c^-) \approx 1.6$ to
$U(T_c^+) \approx 2.0$ is observed. This transition is therefore
first order \cite{compagn}, with a finite latent heat 
$U(T_c^+) - U(T_c^-)$.

At variance with the 1D model, there is no sign of
metastability in the 2D (apart eventually at
$U_c$), at least for what concerns 
the thermodynamical properties. All the data reported
in Figs. \ref{f3} and \ref{f4} correspond to the
time asymptotic state, also in the negative specific
heat region. This is the main not yet
understood difference between the 1D and the 2D model.

\section{Dynamical Properties}\label{subsec:dynamics}

In this Section we discuss the main dynamical properties
of the 1D and 2D models. The most important phenomenon
is correlated to trapping and untrapping of particles within
the cluster in the low energy collapsed phase. this effect
produces an anomalous diffusion (superdiffusion) of particles,
which has a finite time span, being destroyed by 
relaxation to equilibrium in the 1D model and by noisy
fluctuactions (due to the finite number of particles)
of the single-particle potential in the 2D model.
However, for both models in the mean-field limit 
superdiffusion should be present at any time.

\subsection{One dimensional model}\label{subsec:1d_dyn}

\begin{figure}[h]
\centerline{\psfig{figure=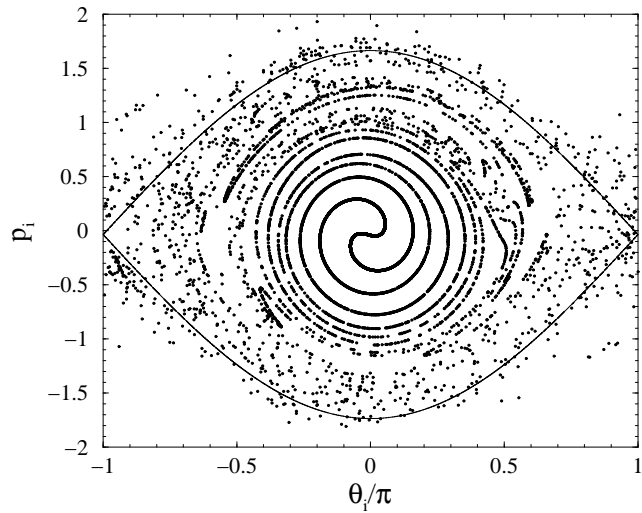,width=7cm,angle=270}}
\caption{Snapshot at some large time of the single-particle
phase-space for an initial "water bag" distribution at $U=0.4$
and with $N=10,000$ for the 1D model. The full line indicates
the separatrix of the unperturbed pendulum drawn at the 
corresponding equilibrium value of $M$.
\label{f5}
}
\end{figure}

\begin{figure}[h]
\centerline{\psfig{figure=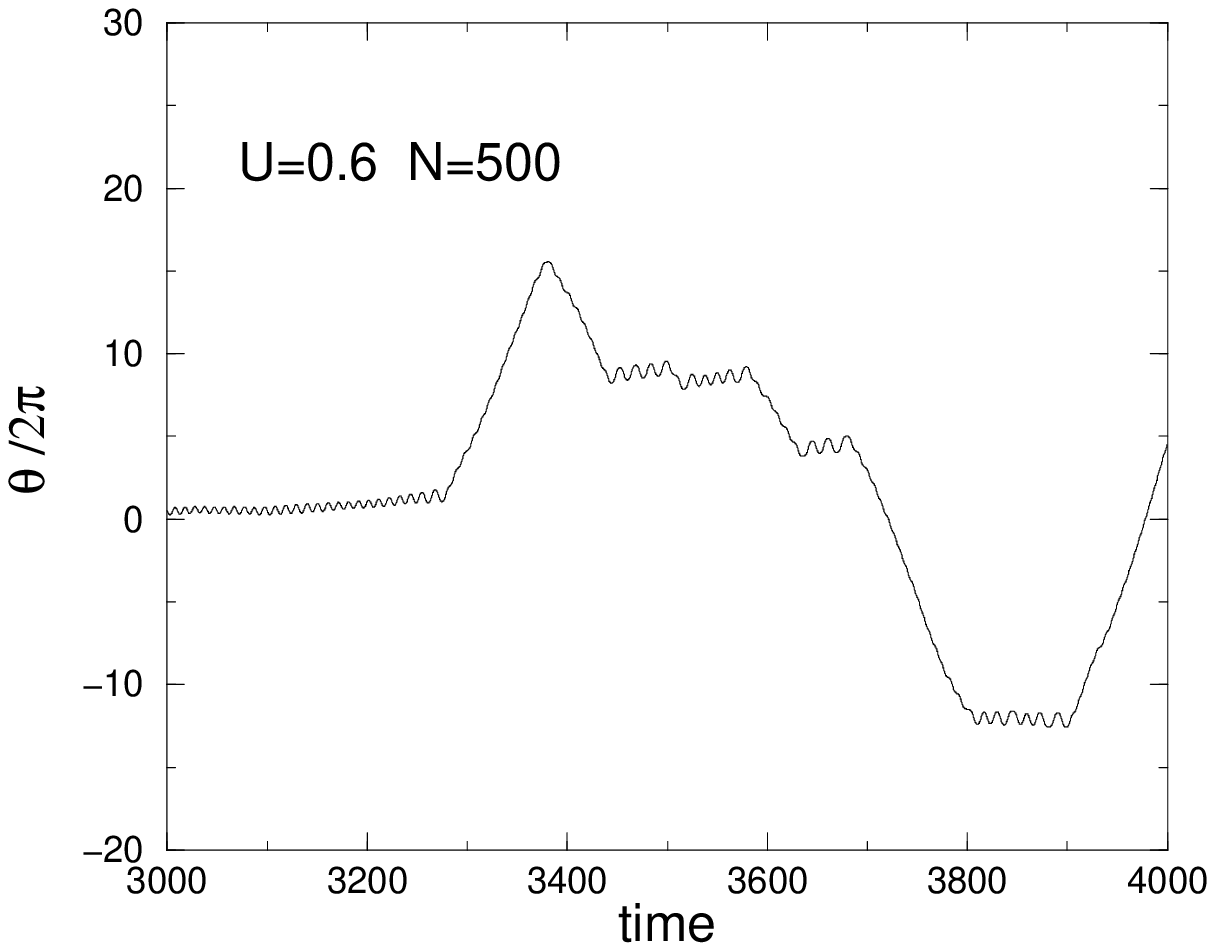,width=11cm}}
\caption{Free flights and trapped motions of
a particle in the 1D model for $U=0.6$, $N=500$
and a "water bag" initial condition.
\label{f7}
}
\end{figure}

The single particle Hamiltonian (5) of the 1D model is that of a
perturbed pendulum. The perturbation has two different sources
in the time dependence of $M$, which generates the stochastic layer
around the separatrix, and in that of $\phi$, which is the main origin
of the trapping/untrapping phenomenon in the nonlinear resonance of
the pendulum. In Fig.~\ref{f5} we show a snapshot taken at some
large time of the particles which were initially put in a "water-bag"
initial condition: near the center of the resonance particles librate
forming a spiral whose branches touch the separatrix region. The
unperturbed separatrix is drawn making reference to the asymptotic
value of $M$ and its center in momentum is slightly displaced from
zero, because the cluster is slowly drifting~\cite{ant1}. Out-of-resonance
particles are capable of performing long flights if energy is large
enough. A typical trajectory of a particle is shown in Fig.~\ref{f7} and 
reveals the presence of tracts of almost free motion with a typical
positive or negative speed, interrupted by tracts where the particle is 
trapped in the cluster, performing with it a much slower motion. It has
been shown in Ref.~\cite{lat2} that the motion is "superdiffusive", i.e.
the mean square displacement $<\theta^2>$ (where the average is
taken over the particles and over distinct time origins) has the
anomalous scaling
\begin{equation}
<\theta^2> \sim t^\alpha
\end{equation}
with $\alpha=1.4 \pm 0.1$ in a given energy range just below the
phase transition point, $0.6 < U < 0.7$, and for a finite time. When the
system relaxes to thermal equilibrium, diffusion becomes normal
($\alpha=1$), but the time taken to do this change increases with $N$,
exactly as the relaxation time to equilibrium. We can then state that in
the thermodynamic limit the motion is superdiffusive forever, when
starting from a non-equilibrium (e.g. "water-bag") initial state. It is
suggestive, and deserves further investigation, the fact that
superdiffusive motion is found in the region of negative specific 
heat, where a core of collapsed trapped particles coexists with a 
"halo" of almost free particles.

\subsection{Two dimensional model}\label{subsec:2d_dyn}

We concentrate in this Subsection on the description of the transport 
properties of the 2D model (\ref{2dham}). 
In the collapsed phase each particle of our system moves
in the single particle potential $V_i=V_i(t)$ (see Fig.~\ref{f2})
of the egg-crate form \cite{geis}.
For a single particle moving in such fixed potential the self dynamics
is known to be anomalously diffusive when the particle is channeling,
{\it i.e}. when its energy lies between $V_M$ and $V_s$ \cite{geis}.
This anomalous behaviour is due to the competition
of laminar and localized phases. Indeed, channeling
particles intermittently show an almost ballistic motion
along the channels of the potential interrupted by localized sequences,
where the particles bounces back and forth on the maxima of the
potential.

Usually anomalous diffusion  has been studied in
systems with a few degrees of freedom \cite{zum2,geis}.
Only few attemps have been made to consider $N$-body
dynamics with $N \gg 1$ \cite{kan1,kan2,kon2}. 

In this Subsection  we give
some indications relative to the basic dynamical mechanisms
governing single particle transport.
To this aim, we consider the time dependence of the mean square
displacement (MSQD), that usually reads as
\begin{equation}
<r^2(t)> \propto t^\alpha
\label{msqd}
\end{equation}
where the average $<.>$ is performed over different time origins and
over all the particles of the system.
The transport is said to be anomalous when $\alpha \ne 1$:
namely, it is subdiffuse if $0 < \alpha < 1$, superdiffusive
if $1 < \alpha < 2$ and ballistic for $\alpha = 2$ \cite{geis,gri1,zum2}.
The usual Einstein diffusion law corresponds to $\alpha=1$
and in 2D can be written as $<r^2(t)> = 4 D t$, where $D$ is the
self-diffusion coefficient.

\begin{figure}[h]
\centerline{\psfig{figure=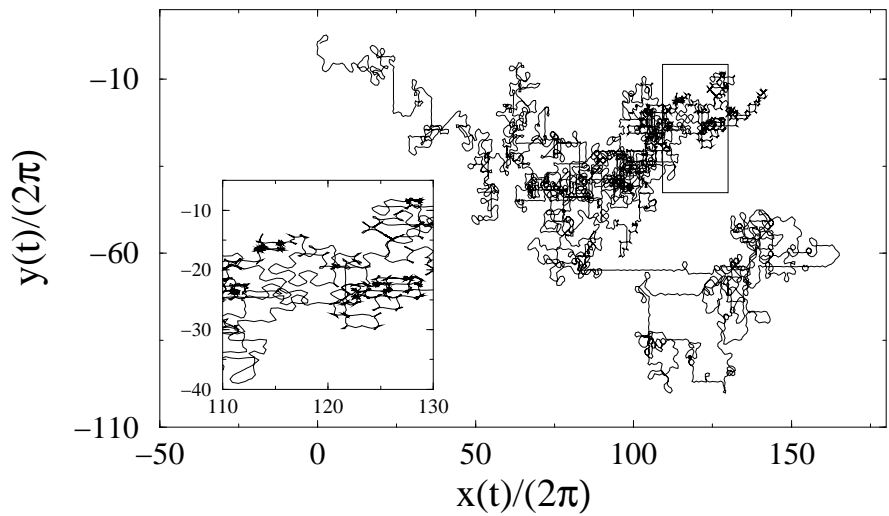,width=7cm,angle=270}}
\caption{Typical time evolution of a particle initially 
in a channel of the 2D single particle potential 
\protect\ref{2dham1}.
In this representation, the $2-D$ torus on which the dynamics takes 
place is unfolded and represented as an infinite plane constituted of 
an infinite number of juxtaposed elementary periodic cell of size 
$1 \times 1$. 
In the inset an enlargement of the trajectory in the indicated box is
reported.  
\label{f8}
}
\end{figure}

If one considers a single particle with an initial energy
between $V_M$ and $V_s$ and follows its trajectory 
for some time
it displays features quite similar to the so-called
L\'evy walks \cite{zum2} (see Fig.\ref{f8}). This kind of
trajectories are usually identified when anomalous diffusion occurs.

As we already mentioned, finite $N$-effects play a relevant role in the
dynamics. Due to self-consistency, they are responsible for the
fluctuations in time of the mean-field quantities $M_{1,2}$ and $P_{1,2}$.
The potential experienced by each particle thus fluctuates
in time and particles having an energy close to $V_s$ have the possibility to
be trapped in the potential as well as to escape from it.
This implies that the localization phenomena illustrated in
Fig.\ref{f8}
are not only due to bounces of the particle on the maxima of the potential,
but also to trapping in the potential well itself due to separatrix crossing.

\begin{figure}[h]
\centerline{\psfig{figure=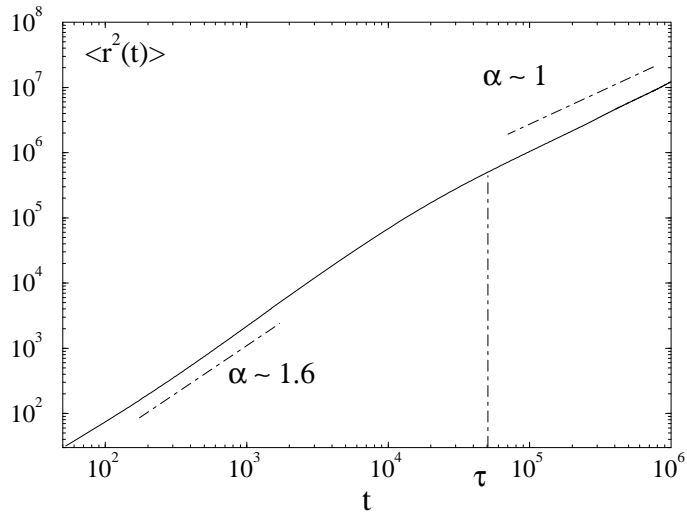,width=7cm,angle=270}}
\caption{Time dependence of the mean square displacement $<r^2(t)>$ 
in a log-log scale for $U=1.1$ and $N=4,000$ for the 2D model.
The solid line indicate the simulation results.
The dot-dashed segments are the estimated slopes 
of $<r^2(t)>$ for $t < \tau$ and $t > \tau$. 
\label{f9}
}
\end{figure}

We have argued from simple considerations that diffusion should be
anomalous in the collapsed phase. This is indeed the case, as confirmed from
direct evaluations of the MSQD in a quite extended energy interval.
An example of this is reported
in  Fig.\ref{f9} for $U=1.1$ and $N=4,000$.
Diffusion is anomalous for times smaller than a crossover time
$\tau$ beyond which the Einstein's diffusion law is recovered
$<r^2(t)> = 4 D t$.  A similar behaviour for the MSQD was previously
observed for a system of $N$
symplectic (globally and locally) coupled maps \cite{kan2,kon2}, but with
a subdiffusive (i.e. with $\alpha < 1$) short time dynamics.

The energy dependence of the $\alpha$-values
is illustrated in Fig.~\ref{f3}.
It shows up clearly that the thermodynamical phase transition
from collapsed to homogeneous phase is associated to a dynamical 
transition from
superdiffusion (with $1.3 < \alpha < 1.9$ for $0.4 \le U < 2.0$)
to ballistic motion (with $\alpha \simeq 2$ for $U \ge U_c$).
In the collapsed regime, we observe an increase of $\alpha$ from
$1.3 \pm 0.1$ to $1.9 \pm 0.1$, that is due to the
modification of the shape of the single particle potential.

A possible explanation of this behaviour is the following.
At very low energy $U < 0.3$, all the particles are essentially
trapped in the potential well and none is channeling: no diffusion
is observed in this case.
For increasing energy a fraction of particles (due to the decrease
of $V_M - V_m$) escapes from the well and some of them
get enough energy to move along the channels: anomalous diffusion
is then evidenced.  The increase
in the value of the exponent $\alpha$ is due to the fact that also
the channels width $V_M-V_s$ grows with $U$.
However, for $U$ approaching $U_c$ the number of untrapped
particles increases noticeably, but now the channel width vanishes
abrubtly. This implies that a significant fraction of particles will move
freely (with energy $ > V_M$). These mechanisms lead naturally to ballistic
motion for $U > U_c$, where the potential $V_i$ is now almost
constant apart from fluctuations of order ${\cal O}(1/\sqrt{N})$.

The fact that in the asymptotic limit $(t \to \infty)$ normal diffusion is
recovered constitutes a typical signature of a noisy dynamics
\cite{flo1,bet1,kan2}. But in our system no external noise is added
to the system, therefore the transition
from anomalous to asymptotic ordinary diffusion must be attributed
to a "deterministic source of noise", that is intrinsic
in our system and due to finite size effects.

A deterministic anomalously diffusing dynamical
system submitted to weak environmental white noise shows
a transition from anomalous diffusion to standard diffusion on
sufficiently long time scales \cite{flo1}.
The crossover time $\tau$ for this transition can be explicitely computed
\cite{flo1} and turns out to increase as an inverse power law of
the noise amplitude, when short time behaviour is superdiffusive.

When analyzing the $N$ dependence of $\tau$ for two energies, 
$U=1.48$ and $U=2.0$, we find that $\tau \propto N$. 
The interpretation of the $N$ dependence
of the crossover time is straighforward if we consider finite $N$ effects
as a source of noise in our model, with a typical amplitude 
of order $1/\sqrt{N}$.
This last assumption is justified by the fact that the microscopic
dynamics of the particles generates stochastic fluctuations
$O(1/\sqrt{N})$ in the values of $M$ and $P$.
These fluctuations become weaker for increasing $N$ and thus naturally
yield an increasing value of $\tau$.
More details about the
relevant dynamical mechanisms can be found elsewhere \cite{tor1}.

We would like to stress that in the mean-field limit the diffusion
will be anomalous at any time, while for finite $N$ normal
diffusion will be always recovered asymptotically.
The fact that the dynamics of the system is
strongly influenced by the order in which the two limits 
$N \to \infty$ and $t \to \infty$ are taken should be
related to the long-range nature of the forces present
in our model.

\section{Chaotic properties}\label{subsec:chaos}

In order to complete the description of the two models,
we investigate in this Section
a fundamental indicator that characterizes the dynamics
of Hamiltonian systems: the maximal Lyapunov exponent $\lambda$.
Our analysis relies on numerical estimations of $\lambda$,
performed considering the evolution in the tangent space
and applying the standard technique first 
introduced in Ref. \cite{benettin}.

Our models are integrable in the limit
of low and high energy, therefore $\lambda \to 0$
both for $U \to 0$ and $U \to \infty$. In between these
two extrema we expect that a finite Lyapunov exponent
will be observed.

\subsection{One dimensional model}\label{subsec:1d_chaos}

\begin{figure}[h]
\centerline{\psfig{figure=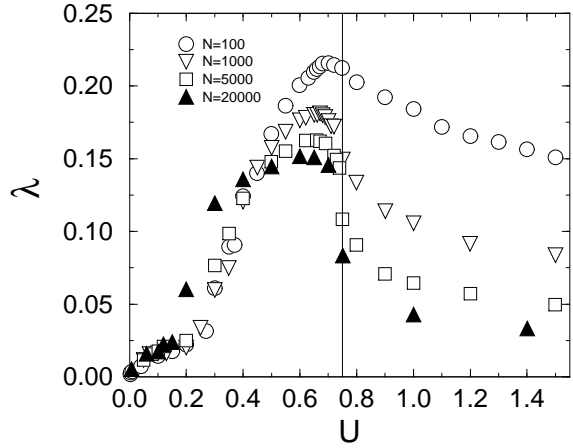,width=11cm}}
\caption{Maximal Lyapunov exponent $\lambda$ 
as a function of the internal
energy for the 1D model (\protect\ref{1dham})
for increasing $N$. The system
has been started from equilibrated initial data.
\label{f10}
}
\end{figure}

\begin{figure}[h]
\centerline{\psfig{figure=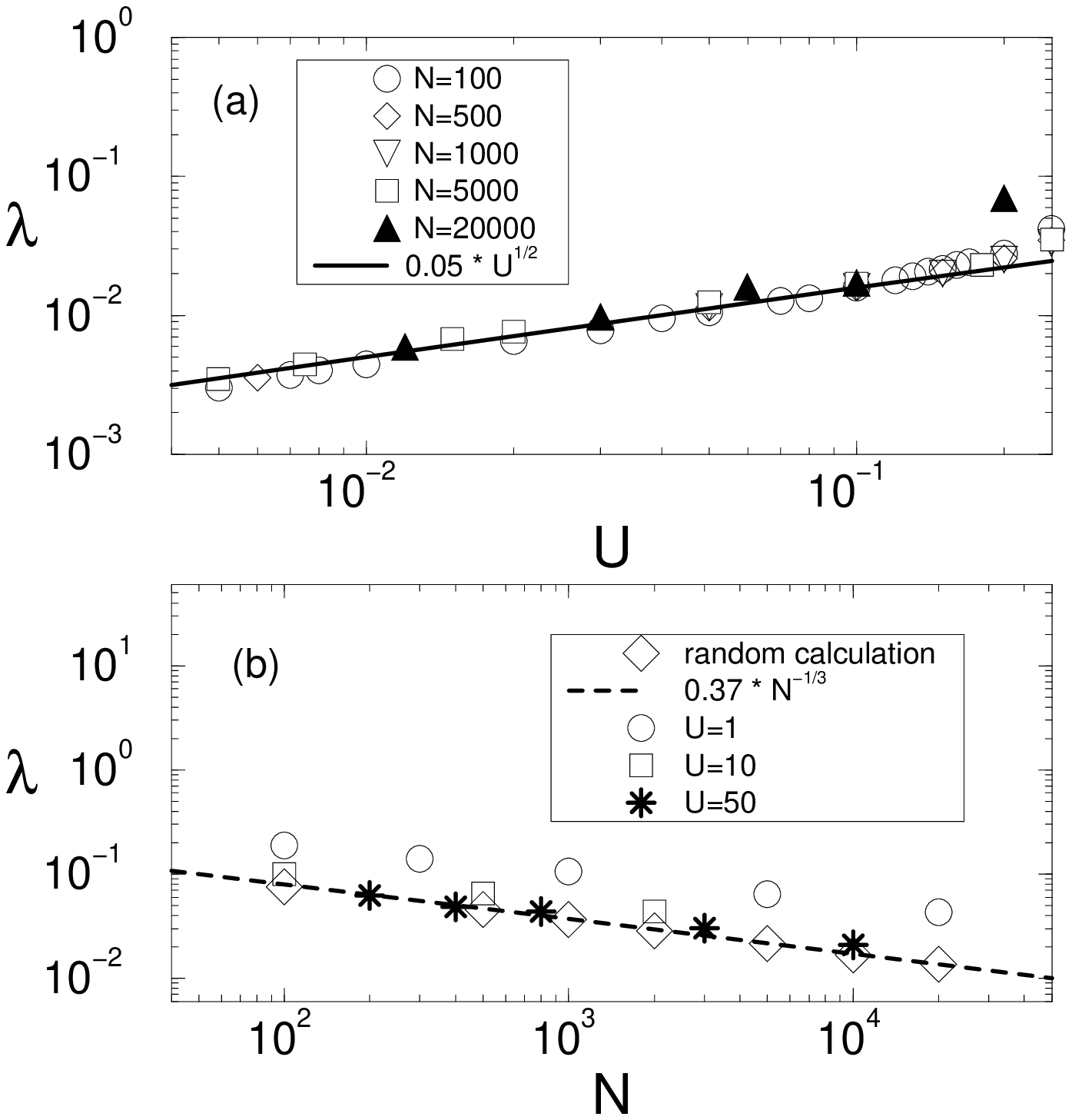,width=11cm}}
\caption{a) Maximal Lyapunov exponent $\lambda$ vs.
the internal energy $U$ for the 1D model
various systems sizes
$N$, showing $N$-independence and the $U^{1/2}$
common scaling.
b) $\lambda$ vs. $N$ for various values of
$ U > U_c = 0.75$. For increasing value sof
$U$ the scaling law $N^{-1/3}$ is verified better
and better. The stars refer to a simulation
performed with random Jacobians, showing the 
effectiveness of the random matrix approximation.
\label{f11}
}
\end{figure}

The 1D model is Lyapunov unstable ($\lambda >0$) over all the
energy range, see Fig.~\ref{f10} (the maximal Lyapunov exponent
of model (\ref{1dham}) was first computed in Ref.~\cite{Yam}).
Below $U\approx 0.2$ there is a weak size dependence of the Lyapunov
exponent and a substantial independence on the chosen initial
condition. A scaling law $\lambda \sim U^{1/2}$ is clearly visible
(see Fig.~\ref{f11}a) and still lacks a convincing theoretical explanation
(the theoretical expression derived in Ref.~\cite{firpo} gives 
$\lambda \sim U$, in sharp disagreement with numerical data); an
argument, reported in Ref.~\cite{lat1} heuristically justifies the
correct scaling. For $U > 0.2$, $\lambda$ rapidly increases and reveals a 
strong dependence on $N$. In fact, it can be shown that in all
the vanishing $M$ phase ($U > U_c$), $\lambda$ vanishes as $N^{-1/3}$
(see Fig.~\ref{f11}b)).
This scaling is easily obtained resorting to a random matrix
approximation of the dynamics~\cite{lat1}, which considers the
successive Jacobians as random and statistically independent, and
has been also derived theoretically in Ref.~\cite{firpo} 
using differential geometric techniques. 
Moreover, also in true one dimensional gravitational models
$\lambda$ vanishes as an inverse power law of $N$ as
shown in \cite{mil1,gouda} (in those cases the exponent
ranges from $1/5$ to $1/4$).  
For true 3D gravitational models Gurzadyan and Savvidy, 
by relating the average Riemaniann curvature to the corresponding 
Lyapunov spectrum, were able to suggest the following
scaling law~\cite{Gurzadyan} :
\begin{equation}
\lambda_G \propto \frac{G M N^{2/3}}{V^{2/3} <v>}
\label{gurz}
\end{equation}
As already reported in Sec. 2.1, our models can
be connected to true gravitational models
by simply changing the time scale by a factor
$1/\sqrt{N}$, since in our model the mass
$M$ is unitary, the gravitational constant 
$G$ is dimensionless, and
the volume $V=(2 \pi)^d$ (where $d$ is the
dimension of the system) do not depend on $N$.
Therefore it is sufficient in Eq.(\ref{gurz})
to rescale the average velocity of the particles
$<v>$ and the Lyapunov exponent by a factor $\sqrt{N}$.
After such rescaling one obtains from Eq.(\ref{gurz})
$\lambda \propto N^{-1/3}$. This suggests
the validity of our results also in the 3D gravitational 
case. The maximal Lyapunov exponent relaxes to
its asympotic value with a $N^{-1/3}$-law also
for a gas of hard-spheres~\cite{posch3}: this 
universality would deserve further investigation.

\noindent

Although the strong $N$ dependence hides the effect, an evident peak of
$\lambda$ is present just below the phase transition point, in 
correspondence of the negative specific heat region. Chaos is "stronger"
in this region due to the contribution of particles which evaporate from
the cluster, entering the stochastic layer near the separatrix. The 
curious coexistence of negative specific heat, superdiffusion and maximal 
chaos deserves further studies.
A confirmation of such a behavior of the maximal Lyapunov exponent
for systems with long-range interactions can be found in
Ref.~\cite{Anteneodo}, where, slightly modifying model (\ref{1dham}),
it has been shown that when the interaction becomes short-ranged $\lambda$
no more vanishes in the $N \to \infty$ limit. These authors
further argue that the exponent $-1/3$ of the infinite-range model
modifies continuously by reducing the range of the interaction, reaching
zero when the interaction becomes short-ranged.

\subsection{Two dimensional model}\label{subsec:2d_chaos}

\begin{figure}[h]
\centerline{\psfig{figure=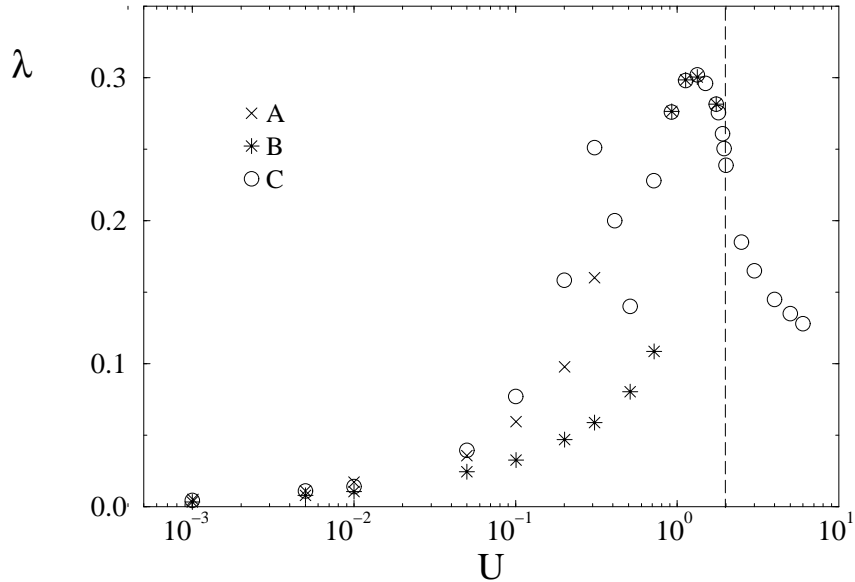,width=8cm,angle=270}}
\caption{Maximal Lyapunov exponent as a function of $U$ in a 
log-linear plot, for three different initializations
of the 2D model. 
The dashed line indicates the critical energy $U_c$.
The reported data corresponds to $N=200$ and to 
integration times ranging from 
$t \sim 1 \times 10^6$ to $t \sim 9 \times 10^6$.
\label{f12}
}
\end{figure}

Our data for the 2D model (\ref{2dham})
are reported in Fig.~\ref{f12} for three
different types of initial conditions :
\begin{itemize}

\item
(A) Maxwellian velocity distribution
and clustered particles;

\item
(B) Maxwellian velocity distribution,
but with a thermal velocity coinciding with its canonical
prediction, and particles spatially organized in a single cluster 
in such a way that also the $M$- and $P$-values reproduce the
canonical prediction;

\item
(C) water-bag velocity distribution and
clustered particles.

\end{itemize}

The initial condition (C) is the one commonly  used through the present
paper, in particular for the study of transport in the system.
As it can be seen from Fig.~\ref{f12} , $\lambda$ grows for
increasing $U$ up to a maximum value and then
decreases. Such maximum (at least for $N=200$) is located
at an energy $U \simeq 1.3-1.4 < U_c$.
For $U > U_c$, we observe a power law decrease of
$\lambda$ with $N$ with an exponent $\sim 0.31$ 
in good agreement with the results for the 1D model.

In our model the vanishing
of $\lambda$ in the mean-field limit, for $U > U_c$, is
connected to the fact that the single particle potential
becomes flat and the system integrable.

In the low energy limit (for $U < 0.01$) a power
law increase of the type $U^{1/2}$ is clearly
observable for all the three types of initial conditions
similarly to 1D. This indicates
that for fully coupled Hamiltonian systems this
property holds in general, independently of the
space dimensionality. In particular, for $U \to 0$
the particles are all clustered, therefore we expect
that the scaling $\lambda \propto U^{1/2}$ should
be related to a "collective" chaotic mechanism.

Let us now discuss the mechanisms
behind the observed behaviours of $\lambda$.
Once the energy
$U$ is fixed for all the three considered initial
conditions, after a reasonable transient, we obtain
exactly the same values for the average temperature
$T$ and magnetizations $M$ and $P$ and a common
Maxwellian distribution for the velocities.
However, at low energies ($0.001 < U < 0.8$) the measured
average $\lambda$ depends heavily on the initial
conditions. This clearly indicates the coexistence
of several equivalent states, that can be considered
as equilibrated within the examined time interval.
Evidently the measurement of $\lambda$ is more sensible
to the presence of such equivalent states than that of
thermodynamical variables.
It should be noticed that this
kind of behaviour is unexpected in $N$-body Hamiltonian
systems, because it is commonly believed that for sufficiently
large values of $N$ Arnold diffusion takes place and each orbit
is allowed to visit the full phase-space. But our
data instead indicate that some "barrier" in the
phase space still survive even for $N =200$.
The origin of this lack of ergodicity is related
to the long range nature of the forces that
induces a persistent memory of the initial
conditions, as previously noticed by Prigogine
and Severne \cite{prigo} for gravitational plasmas.
Recently, some numerical evidence of non ergodicity
has been reported also for 1D gravity \cite{mil1,gouda}. 
But ina recent paper \cite{mak} this evidence is questioned
in favour of a slow relaxation mechanism affecting high
energy particles.

\begin{figure}[h]
\centerline{\psfig{figure=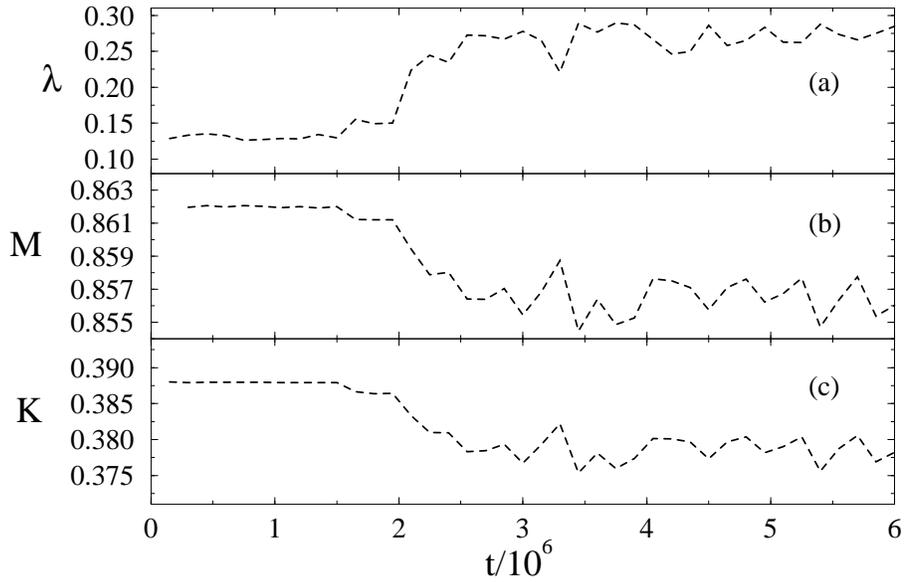,width=8cm,angle=270}}
\caption{Time evolution of the maximal Lyapunov exponent 
$\lambda$, of the magnetization $M$, and of the kinetic 
energy $K$ are shown for a initial condition of (B)-type
for the 2D model at $U=0.87$ and with $N=200$.
\label{f13}
}
\end{figure}

It is clear from Fig.\ref{f12} that, in the
interval $U \in [0.001,0.8]$, for initial conditions
of type (B) $\lambda$ remains always smaller
than the corresponding exponents obtained with
initial conditions (A) and (C).
The maximal differences
are observed in the energy range $0.2 < U < 0.8$,
where particles begin to escape from the cluster
(this happens for initial conditions (A) and (C)).
Above $U \simeq 0.9$ the same Lyapunov is obtained
for all types of initial conditions. A typical feature
of initial conditions (B), for $U < 0.8$ is that
all the particles are trapped in the potential well.
Instead when
one or more particle escape from the cluster ($U > 0.9$)
also with this initial condition the common value of $\lambda$ is
obtained. Indeed, two chaotic mechanisms are present
in the system: one felt by the particles moving
in the minimum of the potential and one by particles
visiting a region near to the separatrix. This second
mechanism is well known and it is related to the chaotic
layer around the separatrix.
The former one is due to the 
erratic motion of the minimum of the potential well.
In order to clearly identify such mechanisms we have followed
the trajectory of a system initially prepared in condition
(B) for an energy $U = 0.87$ and $N=200$. On a short time scale 
all the particles
are trapped and we measure an average value $\lambda \simeq 0.13$.
When at a later time one particle escapes from the
cluster $\lambda$ almost doubles its initial value
(see Fig.~\ref{f13}).
In Fig.~\ref{f13} the magnetization $M$ and the kinetic
energy $K$ are also shown. When the particle escapes
from the cluster $M$ shows a clear decrease as well as $K$.
This last effect is due to the fact that potential energy $V$
is minimal when all the particles are trapped, therefore if
a particle escapes $V$ increases and, due to energy conservation,
$K$ decreases. This is the phenomenon underlying the
negative specific heat effect. From the above arguments
we can identify a "strong" chaos felt from the particles
approaching the separatrix and a "weak" chaos associated
to the orbits trapped in the minimum of the potential.
The presence of these two chaotic mechanisms, together
with the non-ergodicity of the system, explains the
strong dependence of the values of $\lambda$
on the initial conditions.

\section{Conclusions}

This paper reviews, discussing differences and similarities, a class
of models of point masses without hard-cores and
interacting with an infinite-range attractive potential  
in one and two dimensions.
Both models may be seen as toy systems for studying long-range
attractive interactions in an extremely simplified setting. 
We consider them a dynamical extension of the 
lattice gas model introduced and studied 
by Hertel and Thirring~\cite{her2}. 
Thirring~\cite{her1,her2} pointed out that a transition
from a clustered to a homogeneous phase can be associated
with a negative specific heat (within the microcanonical
ensemble) either (I) when the potentials are thermodynamically 
unstable~\cite{her1} or (II) when the thermodynamical potentials
are extensive quantities, but the interaction is long-ranged
(at least in the limit $N \to \infty$).
In a true gravitational potential both the above conditions 
are fulfilled. Studies of $N$-body models 
with short-range  thermodynamically unstable 
potentials~\cite{compagn,posch} 
confirmed that condition (I) was sufficient to observe
transition from a clustered to a gaseous phase, negative
specific heat, inequivalence of canonical and microcanonical
ensemble. Our studies concerns a continuous time model
with thermodynamically stable potential ($H_N \geq E_0 N$), 
but with long-range attractive interaction. Therefore they
confirm that the characteristic associated to the transition 
observed by Hertel and Thirring for a simple lattice gas model 
are present also in more realistic models and that condition
(II) is indeed a sufficient one.
Let us try to summarize condition (I) and (II) in an unique 
prescription : {\it in order to observe a transition belonging 
to the Hertel and Thirring universality class it is necessary 
that the thermodynamical potentials
(e.g. the entropy) are not additive even in the limit 
$N \to \infty$ and that the interaction is 
attractive~\cite{mukamel}}. In other words if 
a system composed by an infinite number of particles
is splitted in two sub-systems typically any thermodynamical
potential
of the whole system coincide with the sum of the entropies
of the two sub-systems, if this is not the case
violation of the standard thermodynamics should be
expected.

Moreover, we show here that extremely
simple infinite-range mean-field models 
display several interesting
features, which are present also in gravity in one and two dimensions
(we believe that our approach can be extended to three dimensions
without conceptual difficulties). A drawback of our models is that the
microcanonical solution has not yet been obtained exactly and we must
rely on numerical simulations (apart from some results which can be
obtained by the collisionless Boltzmann-Poisson equation).
This drawback turns however into an advantage if one is interested in
dynamical effects, about which thermodynamics gives no hint.
The main dynamical effect discussed in this review is "superdiffusion"
(the particles mean square displacement grows faster than linear with
time). Superdiffusion is present in the region of negative specific heat
and is due to the particles which evaporate from the cluster and
perform long flights before being trapped again (the configuration
space is a torus). Evaporation is also the source of "strong" 
chaos: in the negative specific heat region the maximal Lyapunov exponent
$\lambda$ is much larger than everywhere else. Although gravity has
no phase transition in 1D (the gravitational system is always
clustered) and the question is open whether a phase transition
is present in 2D~\cite{abdalla}, both the effects discussed above
could be present in the negative specific heat region of self-gravitating
systems, and we do not see any obstruction to their existence also in
3D. Finally, let us comment on a few scaling laws that we have found
to be universal for all the models of the class we have introduced. 
In the gaseous phase the maximal Lyapunov exponent $\lambda$ vanishes
as $N^{-1/3}$ in analogy with what is found for some specific
gravitational systems and for other dynamical models of 
fluids~\cite{Gurzadyan,posch3}.
In the low energy clustered phase $\lambda \sim U^{1/2}$. This
should be true also in 3D and could be checked in "true" self-gravitating
systems in their collapsed low-energy phase.

\section*{Acknowledgments}\label{sec:concl}

This review is an account of an ongoing joint collaboration of the
authors with V. Latora and A. Rapisarda; many of the results
here reported stem from discussions with them and are part of common
papers. We also thank Andrea Rapisarda for furnishing updated versions
of some figures.
We acknowledge useful discussions with
D. Mukamel, W. Hoover, H.A. Posch, and H. Spohn.
We also are grateful to the DOCS research group 
(see {\it http://docs.de.unifi.it/$\sim$docs})
in Firenze for stimulating interactions.
M.A. thanks Dr. J. Kister for his determinant help during 
his installation in Marseille and J. Albertini for fruitfull
computational help.

\end{document}